\documentclass[
twocolumn,
reprint,
superscriptaddress,
nofootinbib,
nobibnotes,
amsmath,
amssymb,
aps,
prr,
longbibliography
]{revtex4-2}

\usepackage[latin1]{inputenc}
\usepackage{amsmath}
\usepackage{amsfonts}
\usepackage{hyperref}
\usepackage{graphicx}
\usepackage{hyperref}
\usepackage{colortbl}
\usepackage{subfigure}
\usepackage{svg}
\usepackage{chemarr}
\usepackage[normalem]{ulem}

\usepackage[T1]{fontenc}
\usepackage{mathptmx}
\usepackage{bbold}
\usepackage{scalerel}
\usepackage{xcolor}

\usepackage{mathtools}

\newcommand*\ch[1]{\ensuremath{\mathrm{#1}}}
\newcommand{\hyref}[1]{\hyperref[#1]{\ref{#1}}}
\newcommand{\dd}{\mathrm{d}}

\newcommand{\orange}[1]

\begin{document}

\title{Chemically Regulated Conical Channel Synapse for Neuromorphic and Sensing Applications}

\author{T. M. Kamsma}
\thanks{These two authors contributed equally to this work}
\affiliation{Institute for Theoretical Physics, Utrecht University,  Princetonplein 5, 3584 CC Utrecht, The Netherlands}
\affiliation{Mathematical Institute, Utrecht University, Budapestlaan 6, 3584 CD Utrecht, The Netherlands}
\author{M.S. Klop}
\thanks{These two authors contributed equally to this work}
\affiliation{Institute for Theoretical Physics, Utrecht University,  Princetonplein 5, 3584 CC Utrecht, The Netherlands}
\author{W. Q. Boon}
\affiliation{Institute for Theoretical Physics, Utrecht University,  Princetonplein 5, 3584 CC Utrecht, The Netherlands}
\author{C. Spitoni}
\affiliation{Mathematical Institute, Utrecht University, Budapestlaan 6, 3584 CD Utrecht, The Netherlands}
\author{B. Rueckauer}
\affiliation{Department of Artificial Intelligence, Donders Institute for Brain, Cognition and Behaviour, Nijmegen, The Netherlands}
\author{R. van Roij}
\affiliation{Institute for Theoretical Physics, Utrecht University,  Princetonplein 5, 3584 CC Utrecht, The Netherlands}

\date{\today}

\begin{abstract}
Fluidic iontronics offer a unique capability for emulating the chemical processes found in neurons. We extract multiple distinct chemically regulated synaptic features from an experimentally accessible conical microfluidic channel carrying functionalized surface groups, using finite-element calculations of continuum transport equations. By modeling a Langmuir-type surface reaction on the channel wall we couple fast voltage-induced volumetric salt accumulation with a long-term channel surface charge modulation by means of fast charging and slow discharging. These nonlinear charging dynamics emerge across several orders of magnitude of reaction rates and equilibria, and are understood through an analytic approximation rooted in first-principles. We show how short-and long-term potentiation and depression, frequency-dependent plasticity, and chemical-electrical signal spike-timing dependence and coincidence detection (acting like a chemical-electrical AND logic gate), akin to the NMDA mechanism for Hebbian learning in biological synapses, can all be emulated.
\end{abstract}

\maketitle

\section{Introduction}
The pursuit for harnessing the cognitive capabilities of the human brain within artificial systems has garnered significant interest over the last few years \cite{Sangwan2020NeuromorphicMaterials,Zhu2020ADevices,Schuman2017AHardware,Schuman2022OpportunitiesApplications}. Recent advances have propelled fluidic iontronic systems into the spotlight as a platform for achieving this objective \cite{Noy2023FluidDevices,Noy2023NanofluidicSplash,Hou2023LearningNanofluidics,Yu2023BioinspiredComputing,Xie2022PerspectiveApplication}. Various voltage-driven fluidic iontronic systems have lately been considered \cite{Wang2012TransmembraneTransport,Sheng2017TransportingMemristor,Ramirez2021NegativeSolutions,Ramirez2024NeuromorphicSolutions,Kamsma2023UnveilingIontronics,Zhang2024GeometricallySystems,Xu2024NanofluidicMemristors}, paving the way for signaling \cite{Kamsma2023IontronicMemristors,Kamsma2024AdvancedCircuit,Robin2021ModelingSlits} and computational applications \cite{Sabbagh2023DesigningCircuit,Li2023High-PerformanceNano/Microchannels,Emmerich2024NanofluidicSwitches,Kamsma2023Brain-inspiredNanochannels,Liu2024BioinspiredCircuits}. The aqueous medium presents unique opportunities for chemical regulation \cite{Han2022Iontronics:Applications}, on which biological synapses also heavily rely \cite{Xia2005ThePlasticity}.  First steps were recently made towards incorporating chemical regulation alongside electrical signaling for long-term conductance change \cite{Han2023IontronicDissolution,Robin2023Long-termChannels,Wang2024AqueousMembranes}, signal-transduction \cite{Xiong2023NeuromorphicMemristor} and-modulation \cite{Ling2024Single-PoreCombinations}. However, the synergy between chemical and electrical regulation in iontronic neuromorphics remains relatively unexplored and developing easy to fabricate systems that can incorporate multiple chemically regulated synaptic features is of importance.

\begin{figure}[ht]
\centering
     \includegraphics[width=0.45\textwidth]{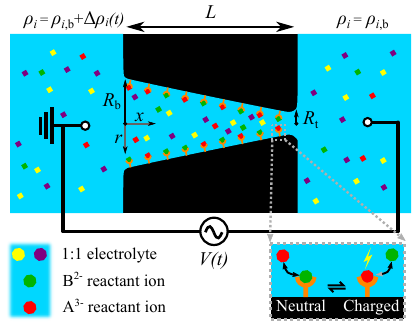}
        \caption{Schematic of an azimuthally symmetric conical channel, carrying functionalized surface groups, of length $L$, base and tip radii $R_{\mathrm{b}}$ and $R_{\mathrm{t}}<R_{\mathrm{b}}$, connecting two reservoirs of an aqueous 1:1 background electrolyte with low concentrations of trivalent and divalent reactant ions A$^{3-}$ and B$^{2-}$. Far from the base and tip the concentrations of ion species $i$ read $\rho_i=\rho_{i,\mathrm{b}}+\Delta\rho_i(t)$ and $\rho_i=\rho_{i,\mathrm{b}}$, respectively. At the base, $\Delta\rho_i(t)$ facilitates optional chemical signals and gradients, alongside electric stimulation via an applied voltage $V(t)$.}
        \label{fig:Fig1}
\end{figure}

\begin{figure*}[ht!]
\centering
     \includegraphics[width=1\textwidth]{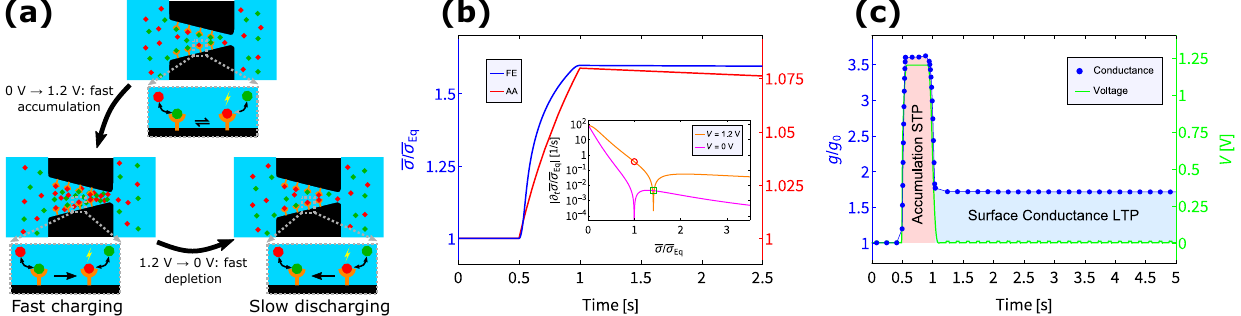}
        \caption{Chemically regulated surface charge and consequent conductance changes upon an 1.2 V pulse for $t\in[0.5,1]$ s. \textbf{(a)} The voltage pulse drives fast ICP, altering reactant ion ratios near the wall, charging the surface, and increasing bulk and surface conductance respectively. After the pulse, salt depletes quickly, while slow surface discharge creates the LTP. \textbf{(b)} Average channel surface charge $\overline{\sigma}$ from full finite-element (FE) calculations and our first-principles analytic approximation (AA), showing qualitative agreement. The inset shows $|\partial_t\overline{\sigma}|$ as a function of $\overline{\sigma}/\sigma_{Eq}$ for $V=0$ V (magenta) and $V=1.2$ V (orange). Charging at $V=1.2$ V (red circle) is orders of magnitude faster than discharging at $V=0$ V (green square). \textbf{(c)} Channel conductance $g(t)/g_0$ from full FE calculations. After the pulse the concentration reverts back to equilibrium quickly, forming the volatile STP, while the surface discharges slowly, driving the LTP.}
        \label{fig:Fig2}
\end{figure*} 

In this work we propose a versatile and tunable chemically regulated platform that features a multitude of synaptic features, based on an experimentally widely available conical channel carrying functionalized surface groups. Through analytic and numerical solutions of Langmuir kinetics and continuum transport equations we couple nonlinear surface charging dynamics \cite{Boon2023CoulombicKinetics} with voltage-induced transient ion concentration polarization (ICP) within the channel interior \cite{Kamsma2023IontronicMemristors}. This ICP facilitates a volatile channel conductance memory \cite{Kamsma2023IontronicMemristors}, analogous to synaptic short-term potentiation (STP) and short-term depression (STD) \cite{abbott2004synaptic,deng2011diverse,rotman2011short}. When coupled to displacement surface charge reactions ICP leads to fast surface charging followed by slow discharging, creating a long lasting conductance increase analogous to synaptic long-term potentiation (LTP) \cite{Abbott2000SynapticBeast}. Neither the charging itself, nor the contrast in (dis)charging rates is necessarily expected, but they elegantly emerge as a consequence of a change in surface potential during ICP and drastically changing concentrations at the surface during and after ion accumulation. The combination of STP and LTP in the same channel enables frequency-dependent plasticity (FDP), a synaptic feature which translates different incoming voltage pulse frequencies into different long-term synaptic strength changes \cite{Kumar2011Frequency-dependentPlasticity,Bliss1993AHippocampus}. Lastly we will turn our attention to chemical signal processing, inspired by the chemical nature of a presynaptic signal \cite{Luscher2012NMDALTP/LTD,Cotman2003ExcitatoryPlasticity,Rauschecker1991MechanismsBeyond,Xia2005ThePlasticity}. Upon an imposed concentration gradient of a reactant ion, a voltage pulse is shown to yield a long-term conductance decrease, emulating long-term depression (LTD) \cite{Abbott2000SynapticBeast}. Conversely, a chemical stimulus in the form of a release of reactant ions at one side of the channel has so significant effect. However, a simultaneous applied voltage and a chemical stimulus yields LTP, thereby forming a coincidence detection mechanism (acting like a chemical-electrical AND logic gate). Moreover, by varying the timing between these signals we retrieve characteristic spike-timing-dependent plasticity (STDP) that uniquely relies on the timing of chemical and electrical stimuli. Interestingly, the coincidence detection and STDP is analogous to the NMDA mechanism \cite{Luscher2012NMDALTP/LTD}, a synaptic mechanism that integrates (presynaptic) chemical and (postsynaptic) electrical signaling dependent on their coincidence and relative timing \cite{Abbott2000SynapticBeast, Bender2006TwoCortex}, believed to be one of the main drivers behind biological LTP and LTD \cite{Luscher2012NMDALTP/LTD,Xia2005ThePlasticity} and consequent Hebbian learning \cite{Cotman2003ExcitatoryPlasticity,Rauschecker1991MechanismsBeyond}. These analogues for neuronal STP, STD, LTP, FDP, LTD, and NMDA-like chemical- and- electric signal STDP and coincidence detection, all materialize in essentially the same artificial and tunable channel.

Conical channels have long been iontronic model systems \cite{yang2019cavity,hu2019ultrasensitive,Jubin2018DramaticNanopores, hou2011biomimetic, ghosal2019solid, bush2020chemical, liu2012surface} due to their nonlinear conductance properties \cite{wei1997current, Boon2022Pressure-sensitiveGeometry, white2008ion, Jubin2018DramaticNanopores, vlassiouk2009biosensing}, which have been extensively studied experimentally \cite{cheng2007rectified, siwy2006ion, bush2020chemical, Jubin2018DramaticNanopores, siwy2002rectification,fulinski2005transport,siwy2005asymmetric}, numerically \cite{duleba2022effect, lan2016voltage, vlassiouk2008nanofluidic, liu2012surface, kubeil2011role}, and analytically \cite{Boon2022Pressure-sensitiveGeometry, dal2019confinement, poggioli2019beyond,uematsu2022analytic}. As a result, cones are now comparatively easy to fabricate  \cite{kovarik2009effect, lin2018voltage, siwy2003preparation, siwy2002fabrication} and are attractive candidates for iontronic (neuromorphic) devices \cite{Powell2011Electric-field-inducedNanopores,Wang2014PhysicalFrequencies,Li2015History-dependentInterfaces,Wang2016DynamicsNanopores,Wang2017CorrelationNanopores,Wang2018HysteresisNanopores,Brown2020DeconvolutionNanopipettes,Brown2021HigherNanopores,Brown2022SelectiveNanopipettes,Ramirez2021NegativeSolutions,Ramirez2023SynapticalMemristors,Ramirez2024NeuromorphicSolutions,Kamsma2023IontronicMemristors,Kamsma2023UnveilingIontronics,Kamsma2024AdvancedCircuit,Liu2024BioinspiredCircuits,Barnaveli2024Pressure-GatedProcessing}. Moreover, conical channels with functionalized surface groups are already widely adopted for sensing application \cite{Song2018UltrasmallLevel,Yi2022GlassEntities,Umehara2009Label-freeProbes,hu2019ultrasensitive,Jia2020TheNanosensor,Voros2021AptamerNanopipettes,Nakatsuka2021SensingNanopipettes,Stuber2023InterfacingTissue}, placing our theoretical work within a clear experimental context. Furthermore, the general principles of coupling transient ICP to nonlinear surface charging can carry over to more specialized devices that feature e.g.\ stronger nonlinear conductance and desirable fabrication properties such as colloid-based \cite{Choi2016HighMembrane,Kamsma2023Brain-inspiredNanochannels,Kim2022AsymmetricDetection} or polyelectrolyte-based \cite{Zhang2024GeometricallySystems} channels.

\section{System and governing equations}
We consider an azimuthally symmetric conical channel, schematically illustrated in Fig.~\ref{fig:Fig1}, with length $L=10\text{\,}\mu\text{m}$, base radius $R_{\mathrm{b}}=120\text{\,nm}$ at $x=0$, tip radius $R_{\mathrm{t}}=30\text{\,nm}$, and channel radius is $R(x)=R_{\mathrm{b}}-x(R_{\mathrm{b}}-R_{\mathrm{t}})/L$ for $x\in\left[0,L\right]$. The channel connects two reservoirs of an incompressible aqueous electrolyte at equal pressure with viscosity $\eta=1.01\text{\,mPa}\cdot\text{s}$, mass density $\rho_{\mathrm{m}}=10^3\text{\,kg}\cdot\text{m}^{-3}$ and electric permittivity $\epsilon=0.71\text{\,nF}\cdot\text{m}^{-1}$. The electrolyte contains four types of ions with equal diffusion coefficients $D=1.5\text{\,}\mu\text{m}^2\text{ms}^{-1}$, mostly ions of valency $z_\pm=\pm1$ with concentrations $\rho_{\pm}(x,r,t)$, but also reactant ions A$^{3-}$ and B$^{2-}$ of valency $z_{\ch{A}}=-3$ and $z_{\ch{B}}=-2$ with concentrations $\rho_{\ch{A(B)}}(x,r,t)$. At the far side of both reservoirs we impose bulk ion concentrations $\rho_{i}(t)=\rho_{i,\mathrm{b}}+\Delta\rho_{i}(t)$, with $\rho_{\pm,\mathrm{b}}=1.01\pm0.05\text{\,mM}$, and $\rho_{\mathrm{A (B),b}}=0.02\text{\,mM}$, such that $\rho_{\mathrm{A(B),b}}\ll\rho_{\pm,\mathrm{b}}$ and overall charge neutrality in the bulk is ensured. The optional $\Delta\rho_{i}(t)$ coarsely emulates an injection of ions into the base reservoir, creating a local source of ions, that consequently dissipate into the reservoir after the influx stops. In Sec.~\ref{sec:NMDA} and in the Supplemental Material (SM) \cite{SM} we further support and show with additional FE results respectively that this forms an effective and sufficiently accurate implementation of a chemical signal.

On the far side of the tip and base reservoir we impose an electric potential $V(t)$ and ground respectively, which leads to an electric potential profile $\Psi(x,r,t)$, an electro-osmotic fluid flow with velocity field $\mathbf{u}(x,r,t)$, and ionic fluxes $\mathbf{j}_{i}(x,r,t)$. The channel wall carries functionalized surface groups with density $\Gamma=1\text{\,nm}^{-2}$ to which ions $\text{A}^{3-}$ (or $\text{B}^{2-}$) can bind to form a charged (or neutral) group $\text{SA}^{-}$ (or $\text{SB}$), such that the area density of charged and neutral groups is $\sigma$ and $\Gamma-\sigma$, respectively, resulting in a negative surface charge density $-e\sigma(x,t)$.

Here we study anionic displacement surface reactions, but all results would identically emerge for cationic reactant ions because of invariance under a complete switch of charge and voltage signs. Divalent-trivalent displacement reactions are well-established for various application \cite{Reichle1986SynthesisHydrotalcite}, though mostly in the context of various other applications \cite{Bukhtiyarova2019AHydroxides} such as e.g.\ heavy metal removal from water \cite{Feng2022AComposites}. Additionally, while displacement reactions yield the largest contrast in (dis)charging timescales, more conventional adsorption reactions also feature the relevant nonlinear charging dynamics \cite{Boon2023CoulombicKinetics}. We calculate $\sigma(x,t)$ self-consistently by accounting for a displacement reaction of the form
\begin{align}
    \ch{SB}_{\ch{(s)}}+\ch{A}_{\ch{(aq)}}^{3-}\xrightleftharpoons[kK\rho_{\ch{B}}]{k\rho_{\ch{A}}}\ch{SA}_{\ch{(s)}}^{-}+\ch{B}_{\ch{(aq)}}^{2-},
\end{align}
where a trivalent ion (for example phosphate or citrate for anionic reactions, or magnesium(II) \cite{Reichle1986SynthesisHydrotalcite,Feng2022AComposites} for cationic reactions) displaces a divalent ion (for example sulfate or carbonate for anionic reactions, or iron(III) \cite{Reichle1986SynthesisHydrotalcite,Feng2022AComposites} for cationic reactions) from the channel wall. The nonlinear kinetics of displacement reactions allows fast charging of the surface to be combined with very slow discharging \cite{Boon2023CoulombicKinetics}. We describe the surface charge density $-e\sigma(x,t)$ by local Langmuir kinetics of the form \cite{Boon2023CoulombicKinetics,Werkhoven2018Flow-InducedKinetics}
\begin{align}\label{eq:langmuirdisplacement}
    \dfrac{\partial \sigma}{\partial t}=k\Big(\rho_{\ch{A}}(\Gamma-\sigma)-K\rho_{\ch{B}}\sigma\Big),
\end{align}
with reaction rate $k=200\text{\,mM}^{-1}\text{s}^{-1}$, equilibrium constant $K=1.25$, and concentrations at the channel wall $\rho_{\ch{A(B)}}=\rho_{\ch{A(B)}}(x,R(x),t)$. The specific reaction parameters $k$ and $K$ are not essential for our results, as we will explain below and show in the SM \cite{SM}, where we also show similar results for a monovalent-divalent displacement reaction. At rest the surface potential is $\psi_{0}\approx -77$ mV for our parameters.

For the interior of the channel and the reservoirs, transport is described by the Poisson-Nernst-Planck-Stokes Eqs.~(\ref{eq:poisson})-(\ref{eq:stokes}) given by 
\begin{gather}
	\nabla^2\Psi=-\frac{e}{\epsilon}\sum_{i}z_i\rho_i,\label{eq:poisson}\\
	\dfrac{\partial\rho_{i}}{\partial t}+\nabla\cdot\mathbf{j}_{i}=0,\label{eq:ce}\\
 \mathbf{j}_{\pm}=-D\left(\nabla\rho_{i}+z_i\rho_{i}\frac{e\nabla \Psi}{k_{\mathrm{B}}T}\right)+\mathbf{u}\rho_{i},\label{eq:NP}\\
	\rho_{\mathrm{m}}\dfrac{\partial\mathbf{u}}{\partial t}=\eta\nabla^2\mathbf{u}-\nabla P-e\nabla \Psi\sum_{i}z_i\rho_i;\qquad\nabla\cdot\mathbf{u}=0.\label{eq:stokes}
\end{gather}
Electrostatics is accounted for by the Poisson Eq.~(\ref{eq:poisson}), the conservation of ions by the continuity Eq.~(\ref{eq:ce}), Fickian diffusion, Ohmic conduction, and Stokesian advection by the Nernst-Planck Eq.~(\ref{eq:NP}), and the force balance on the (incompressible) fluid by the Stokes Eq.~(\ref{eq:stokes}) that includes an electric body force. On the channel wall we impose a no-slip boundary condition $\mathbf{u}=0$, Gauss' law $\mathbf{n}\cdot\nabla\Psi=e\sigma(x,t)/\epsilon$ with $\mathbf{n}$ the wall's inward normal vector, and $\mathbf{n}\cdot\mathbf{j}_{i}=R_i$, where $R_i$ is the surface reaction rate given by $R_{\mathrm{B}}=-R_{\mathrm{A}}=\partial_t\sigma$ and $R_{\pm}=0$. 

\section{Nonlinear surface charge dynamics for long-term potentiation}
We numerically solve the full set of Eqs.~(\ref{eq:langmuirdisplacement})-(\ref{eq:stokes}) with finite-element (FE) calculations using COMSOL \cite{multiphysics1998introduction}. For insights into the mechanism we also derive an Analytic Approximation (AA) by assuming that $\rho_{\mathrm{A}}$ and $\rho_{\mathrm{B}}$ on the channel wall satisfy Boltzmann distributions $\rho_{\mathrm{A(B)}}\propto\exp[-z_{\mathrm{A(B)}}e\psi_{0}(x)/k_{\mathrm{B}}T]$ and are subject to the same (transient) voltage-dependent ICP, i.e.\ $\rho_{\ch{A (B)}}=\rho_{\ch{A (B)}}(V,\psi_0)$ \cite{Boon2022Pressure-sensitiveGeometry,Kamsma2023IontronicMemristors,Kamsma2023UnveilingIontronics} (full details in SM \cite{SM}). As schematically depicted in Fig.~\ref{fig:Fig2}(a), a positive voltage pulse of duration $\Delta T=0.5$ s over the channel induces ion accumulation inside the channel \cite{Boon2022Pressure-sensitiveGeometry} over a timescale $\tau=L^2/12D\approx 5.6$ ms \cite{Kamsma2023IontronicMemristors}, which renders the surface potential $\psi_{0}(x)$ less negative and therefore shifts the concentration ratio at the channel wall $\rho_{\mathrm{A}}/\rho_{\mathrm{B}}=\exp[e\psi_{0}(x)/k_{\mathrm{B}}T]$ towards $\text{A}^{3-}$, thereby charging the surface at a characteristic rate $k\rho_{\ch{A}}\sim\mathcal{O}(1)\text{s}^{-1}$. Upon removal of the applied voltage the salt concentration quickly decreases over the timescale $\tau$, while the extra charge on the surface repels the $\text{B}^{2-}$ ions necessary for the discharging reaction \cite{Boon2023CoulombicKinetics}, both resulting in a significantly slower discharging rate of characteristic time $kK\rho_{\ch{B}}\sim\mathcal{O}(0.01)\text{s}^{-1}$. Therefore we identify ICP (increasing $k\rho_{\ch{A}}$) and Coulombic ion-wall repulsion (decreasing $kK\rho_{\ch{B}}$) as origins of the fast (slow) (dis)charging, irrespective of $k$ and $K$. FE calculations support this insight (see SM \cite{SM}) as we recover the charging-discharging asymmetry across at least three orders of magnitude of $k$ and $K$. Constraints on other parameters are also mild $k\rho_{\ch{A}}\Delta T\sim\mathcal{O}(1)$, $\tau/\Delta T\ll 1$, and the surface conductance contribution needs to be sufficiently high so $4\lambda_{\ch{D}}/(R_{\ch{b}}+R_{\ch{t}})\sim\mathcal{O}(1)$ \cite{Aarts2022IonMicropores} with $\lambda_{\ch{D}}$ the Debye length. Since $R_{\ch{b(t)}}$, $L$ (for $\tau$), $\rho_{\pm,\ch{b}}$ (for $\lambda_{\ch{D}}$), $\rho_{A(B),\ch{b}}$ (for (dis)charging rates), and $\Delta T$ are all experimentally tunable, while the fast (slow) (dis)charging is largely independent from the non-tunable $k$ and $K$, we predict that our results can emerge for a wide range of given anionic or cationic displacement reactions.
\begin{figure}[t]
\centering
     \includegraphics[width=0.49\textwidth]{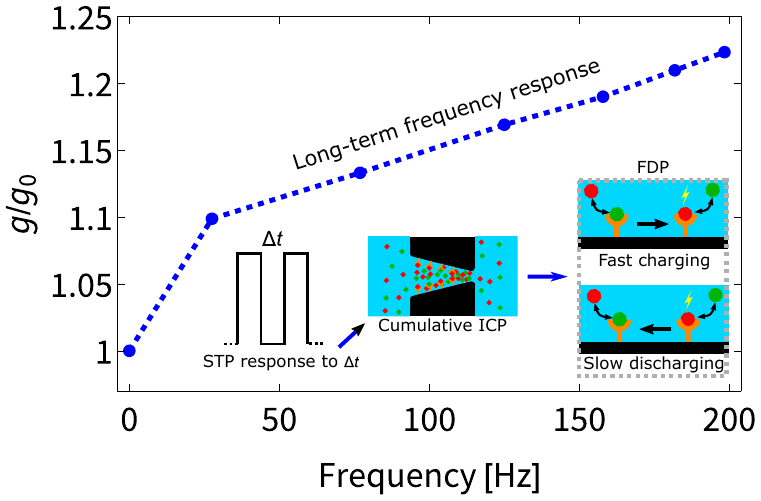}
        \caption{Channel conductance $g/g_0$ after 40 pulses of 0.8 V with fixed duration 3 ms and varying intervals. Salt accumulation is cumulative when pulses are closely spaced, leading to frequency-dependent charging of the wall and consequent LTP (as shown in Fig.~\ref{fig:Fig2}(c)), thereby creating FDP.}
        \label{fig:Fig3}
\end{figure}

\begin{figure*}[t]
\centering
     \includegraphics[width=1\textwidth]{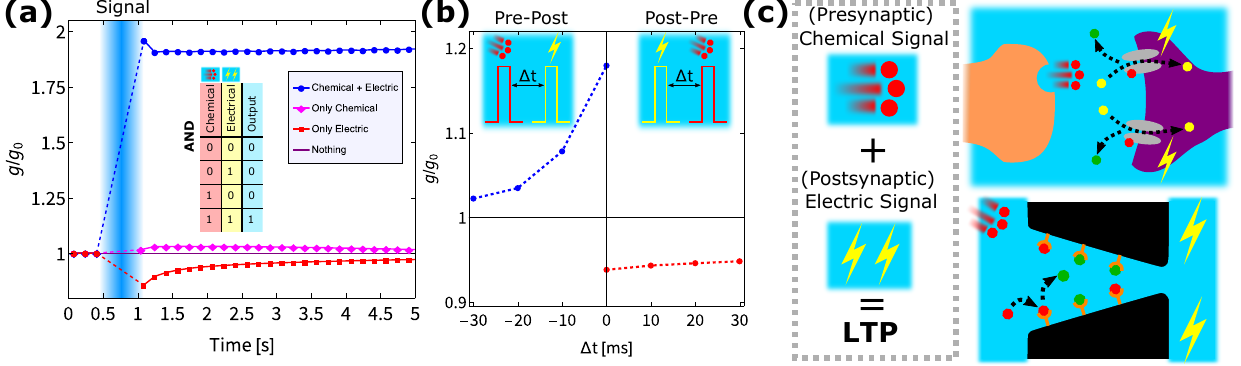}
        \caption{\textbf{(a)} Channel conductance $g/g_0$ response to chemical and/or electric signals. A voltage pulse induces LTD, while a chemical signal, i.e\ temporarily increasing the $\text{A}^{3-}$ base reservoir concentration from $4\cdot10^{-4}$ mM to $0.01$ mM, has a comparatively insignificant effect (magenta). Combined voltage and chemical signals significantly enhance conductance, resulting in LTP (blue). \textbf{(a, inset)} Response in (a) corresponds to a chemical-electrical AND logic gate. \textbf{(b)} Conductance response to chemical and electrical 50 ms signals separated by $\Delta t$ revealing STDP \cite{Feldman2000Timing-BasedCortex,Abbott2000SynapticBeast,Bender2006TwoCortex}. \textbf{(c)} Schematic of the NMDA coincidence detection mechanism \cite{Luscher2012NMDALTP/LTD,Cotman2003ExcitatoryPlasticity,Rauschecker1991MechanismsBeyond,Xia2005ThePlasticity}. NMDA receptors (gray) open if glutamate (red) released by the presynaptic neuron binds to them and $\text{Mg}^{2+}$ (green) is removed by postsynaptic depolarization, allowing a $\text{Ca}^{2+}$ (yellow) influx into the postsynaptic neuron and instigate LTP \cite{Luscher2012NMDALTP/LTD}. Analogously, a simultaneous release of reactant ions at the base and electric signal at the tip increases channel conductance.}
        \label{fig:Fig4}
\end{figure*}
In Fig.~\ref{fig:Fig2}(b) we show the time trace of the average surface charge $\overline{\sigma}(t)=\int_0^L R(x)\sigma(x)\dd x/\int_0^L R(x)\dd x$ of the channel before, during, and after an applied voltage pulse of 1.2 V for $t\in[0.5,1]\text{s}$ as predicted by our FE calculations (blue) and our AA (red). A fast increase of $\overline{\sigma}$ is visible during the voltage pulse, after which $\overline{\sigma}$ decays very slowly. In the inset of Fig.~\ref{fig:Fig2}(b) we plot $|\partial_t\overline{\sigma}|$, as determined by the AA of Eq.~(\ref{eq:langmuirdisplacement}), at an applied 0 V (magenta) and 1.2 V (orange). The initial charging rate at $V=1.2$ V (red circle) is orders of magnitude faster than the initial discharging rate at $V=0$ V (green square). The long-lasting increased surface charge causes a long-term enhanced conductance $g(t)$ shown in Fig.~\ref{fig:Fig2}(c), where $g_0$ is the channel conductance before the voltage pulse. During the voltage pulse the conductance increases to $\sim3.6g_0$ due to the fast ion accumulation, which facilitates STP \cite{Kamsma2023IontronicMemristors}, while a long-term conductance of $\sim1.7g_0$ remains after the pulse due to the slow discharging of the channel surface \cite{Boon2023CoulombicKinetics}, resulting in an LTP that is up to an order of magnitude higher than the chemically induced LTP recently reported in two-dimensional fluidic devices \cite{Robin2023Long-termChannels,Wang2024AqueousMembranes}. The initial decay of $g(t)$ by $0.1-0.2$\%$/\text{s}$ corresponds to a memory retention of order $\sim10$ min, thus even longer than the AA prediction.

\section{Combining STP and LTP for FDP}
Next we combine the STP and LTP features shown in Fig.~\ref{fig:Fig2}(c) to extract FDP. Volatile (fluidic) memristors that exhibit STP are well-known to act as high-pass filters \cite{Xiong2023NeuromorphicMemristor,Kamsma2023Brain-inspiredNanochannels,Zhou2022VolatileComputing,Shahsavari2016CombiningNetworks}. However, their frequency response dissipates quickly after the spike train, while biological FDP is long-term \cite{Kumar2011Frequency-dependentPlasticity,Bliss1993AHippocampus}. Here the frequency-dependent cumulative ICP that responds to the short intervals is converted to a long-term conductance change through the mechanism identified in Fig.~\ref{fig:Fig2}, thereby achieving FDP. Investigating FDP is computationally expensive, because it involves pulse trains rather than a single pulse. To help speed up calculations we changed the parameters (only here) to $R_{\mathrm{b}}=80$ nm, $R_{\mathrm{t}}=20$ nm, $k=800\text{ mM}^{-1}\text{s}^{-1}$, $\rho_{\pm,\mathrm{b}}=1.75\pm0.25\text{ mM}$, and $\rho_{\mathrm{A(B),b}}=0.1\text{ mM}$. We apply a train of 0.8 V pulses of fixed duration 3 ms (i.e.\ $\approx \tau/2$) and variable interval times between pulses to create trains of different frequencies. In Fig.~\ref{fig:Fig3} the normalized conductances are shown after 40 pulses, i.e.\ after an equal cumulative time of an applied voltage for each frequency. Higher frequencies result in a larger conductance increase, which by virtue of the LTP features shown in Fig.~\ref{fig:Fig2}(c) lasts long-term after the spike train has passed. Therefore, though the frequency dependence of neurons can be more sophisticated \cite{Kumar2011Frequency-dependentPlasticity,Bliss1993AHippocampus}, we do achieve the neuronal characteristic that a chemically regulated process facilitates a long-term response to an incoming frequency, thereby creating an analogue to neuronal FDP.

\section{NMDA-like chemical-electrical coincidence and spike-timing detection}\label{sec:NMDA}
Lastly, we extend to chemical signal processing and demonstrate a chemical and electric signal coincidence and timing detection, analogous to the pivotal NMDA coincidence detection mechanism that helps govern our biological synapses  \cite{Luscher2012NMDALTP/LTD,Cotman2003ExcitatoryPlasticity,Rauschecker1991MechanismsBeyond,Xia2005ThePlasticity}. If we apply a concentration gradient of ion $\text{A}^{3-}$ by lowering the base reservoir concentration from the default 0.02 mM to $4\cdot10^{-4}$ mM (adjusting $\Delta\rho_{+}$ for charge neutrality), then we find that a 1.2 V voltage pulse (for $t\in[0.5,1]$\,s) shifts the ratio at the channel wall $\rho_{\mathrm{A}}/\rho_{\mathrm{B}}$ towards ion $\text{B}^{2-}$, discharging the surface. Fig.~\ref{fig:Fig4}(a, red) shows the resulting decreased conductance, which lasts for a considerably longer time than the pulse duration, i.e.\ we find LTD. If we combine the voltage pulse with a concentration shock by increasing $\rho_{\ch{A,b}}$ in the base reservoir  (10 $\mu$m away from the base) from $4\cdot10^{-4}$ mM to $0.01$ mM during $t\in[0.5,1]$ s, then the signals compound to extract the strong LTP shown in Fig.~\ref{fig:Fig4}(a, blue). Solely a chemical shock without a voltage pulse results in comparatively insignificant effect as shown in Fig.~\ref{fig:Fig4}(a, magenta). Interestingly, the results of Fig.~\ref{fig:Fig4}(a) could also be interpreted (see inset) as a chemo-electric logic AND gate. This chemical (electrical) pulse is analogous to the chemical (electrical) signal of a presynaptic (postsynaptic) neuron \cite{Luscher2012NMDALTP/LTD}, where not just the coincidence, but also the relative timing is of importance \cite{Abbott2000SynapticBeast}. We sequentially apply 50 ms signals, separated by a time interval $\Delta t$ that is either negative (first chemical then electric) or positive (first electric then chemical), to a channel with radii $R_{\ch{b} (\ch{t})}/2$ (thinner channels decrease the effect of a sole chemical pulse). We observe characteristic STDP \cite{Abbott2000SynapticBeast,Bender2006TwoCortex} shown in Fig.~\ref{fig:Fig4}(b). For $\Delta t<0$, the extra A$^{3-}$ ions transiently clear after the chemical pulse, creating a diminishing LTP effect for more negative $\Delta t$. 

Increasing the concentration boundary condition at the far end of the base reservoir is a coarse but effective and simple representation of an injection of $\text{A}^{3-}$ at the base side. This approach implements a temporary increase of $\rho_{\text{A}}$ due to the diffusion of $\text{A}^{3-}$ from the injection point through the base reservoir to the channel. This diminishes after the injection stops as the ions in the modeled reservoir dissipate due to the reverted boundary condition. A more detailed implementation would inject the ions near the base and simultaneously keep the boundary concentration fixed further away, which yields essentially the same effects. Indeed, in the SM \cite{SM} we show that the time-dependent potentiation also occurs when $\text{A}^{3-}$ is temporarily injected with an inward flux, while all boundary concentrations are kept constant further away.

For $\Delta t>0$ the voltage pulse induces LTD, but a sole subsequent release of ions has no significant effect. As in Fig.~\ref{fig:Fig4}(a), the LTD response decays significantly in the first several $\sim100$ ms. Therefore, similar to various biological neurons, the time-window for LTD is considerably wider than for LTP \cite{Feldman2000Timing-BasedCortex,Abbott2000SynapticBeast,Bender2006TwoCortex}. However, a limitation is that some LTD response still remains for even longer interval times.

Figs.~\ref{fig:Fig4}(a) and (b) display a fascinating analogy to the biological NMDA chemical and electrical coincidence detection mechanism, schematically depicted in Fig.~\ref{fig:Fig4}(c). This synaptic mechanism integrates (presynaptic) chemical and (postsynaptic) electrical signals to detect the coincidence and timing of pre- and postsynaptic activity, one of the main drivers behind LTP, LTD, and consequent Hebbian learning \cite{Luscher2012NMDALTP/LTD,Cotman2003ExcitatoryPlasticity,Rauschecker1991MechanismsBeyond,Xia2005ThePlasticity,Abbott2000SynapticBeast,Bender2006TwoCortex}. The channel's coincidence detection and STDP are unique as they are not based on voltage spike timing, a common feature of memristors \cite{Schuman2022OpportunitiesApplications,Zhu2020ADevices,Schuman2017AHardware}, but rather rely on the biologically inspired coincidence of a chemical and electrical signal.

\section{Conclusion}
In summary, on the basis of Langmuir kinetics and transport equations for water and dissolved ions, we predict that for a wide range of reaction parameters a rich array of chemically regulated synaptic features emerges from an experimentally accessible conical channel with functionalized surface groups, a type of channel widely adopted for (bio)sensing purposes \cite{Yi2022GlassEntities,Umehara2009Label-freeProbes,hu2019ultrasensitive,Jia2020TheNanosensor,Voros2021AptamerNanopipettes,Nakatsuka2021SensingNanopipettes,Stuber2023InterfacingTissue,Song2018UltrasmallLevel}. These features include short- and long-term potentiation and depression, frequency-dependent plasticity, and chemically regulated coincidence and spike-timing detection, akin to the pivotal NMDA mechanism \cite{Luscher2012NMDALTP/LTD,Cotman2003ExcitatoryPlasticity,Rauschecker1991MechanismsBeyond,Xia2005ThePlasticity,Abbott2000SynapticBeast,Bender2006TwoCortex}. Looking ahead, these channels' coupled concentration polarization with pressure-driven flows offers a potential additional input dimension \cite{Jubin2018DramaticNanopores, Boon2022Pressure-sensitiveGeometry, Barnaveli2024Pressure-GatedProcessing}. Our analytical insights and predicted synaptic features offer a theoretical foundation for chemically regulated neuromorphic iontronics and provide guidance for bringing it into an experimentally accessible domain.

%

\end{document}


\title{Supplemental Material for: Chemically Regulated Conical Channel Synapse for Neuromorphic and Sensing Applications}

\author{T. M. Kamsma}
\thanks{These two authors contributed equally to this work}
\affiliation{Institute for Theoretical Physics, Utrecht University,  Princetonplein 5, 3584 CC Utrecht, The Netherlands}
\affiliation{Mathematical Institute, Utrecht University, Budapestlaan 6, 3584 CD Utrecht, The Netherlands}
\author{M. Klop}
\thanks{These two authors contributed equally to this work}
\affiliation{Institute for Theoretical Physics, Utrecht University,  Princetonplein 5, 3584 CC Utrecht, The Netherlands}
\author{W. Q. Boon}
\affiliation{Institute for Theoretical Physics, Utrecht University,  Princetonplein 5, 3584 CC Utrecht, The Netherlands}
\author{C. Spitoni}
\affiliation{Mathematical Institute, Utrecht University, Budapestlaan 6, 3584 CD Utrecht, The Netherlands}
\author{B. Rueckauer}
\affiliation{Department of Artificial Intelligence, Donders Institute for Brain, Cognition and Behaviour, Nijmegen, The Netherlands}
\author{R. van Roij}
\affiliation{Institute for Theoretical Physics, Utrecht University,  Princetonplein 5, 3584 CC Utrecht, The Netherlands}

\maketitle

\section{Analytic Approximation}\label{sec:AA}
To better understand the features of the mechanism that underpins the long term-potentiation (LTP) presented in the main text, we derive an analytic approximation (AA) that is based on the continuum ion transport equations of the main text. Conical channels with a negative surface charge exhibit accumulation (depletion) of the total ion concentration upon a positive $V>0$ (negative $V<0$) applied voltage over the channel \cite{Boon2022Pressure-sensitiveGeometry,Kamsma2023UnveilingIontronics}, which is analytically described by
\begin{equation}\label{eq:rhos}
\begin{split}
    \overline{\rho}_s(x,V)=2\rho_b-\frac{1}{\text{Pe}(V)/V}\frac{2e \left(\sigma(0)\Delta R+\sigma^{\prime} R_{\mathrm{b}}\right)}{k_{\mathrm{B}}T  R_{\mathrm{t}}^2}\\
    \left(\frac{R_{\mathrm{b}}(1-x/L)}{R(x)}-\frac{e^{-\text{Pe}(V)\frac{(1-x/L)R_{\mathrm{t}}}{R(x)}}-1}{e^{-\text{Pe}(V)\frac{R_{\mathrm{t}}}{R_{\mathrm{b}}}}-1}\right),
    \end{split}
\end{equation}
with $\overline{\rho}_{\mathrm{s}}(x,V)$ the radially averaged total ion (i.e.\ salt) concentration profile, which we use as a proxy for the salt concentration at the channel's central axis. It accounts for the accumulation of salt in the channel upon applying a positive voltage $V>0$. Here $\sigma(0)$ is the surface charge density at the base of the channel and $\sigma^{\prime}$ is a convenient proxy (see below) of the surface charge density difference between tip and base. Note that Eq.~(\ref{eq:rhos}) therefore implies that a spatially inhomogenous salt concentration either requires a conical channel geometry (with $\Delta R\equiv R_{\mathrm{b}}-R_{\mathrm{t}}\neq 0$) or a spatially varying surface charge (with $\sigma'\neq 0$), or both \cite{Kamsma2023UnveilingIontronics}. We also introduced the P\'{e}clet number at the narrow end of the channel $\text{Pe}(V)=Q(V)L/(\pi DR_{\ch{t}}^2)$ with $Q(V)=\pi R_{\mathrm{t}}R_{\mathrm{b}}\epsilon\overline{\psi}_{0} V/(\eta L)$ the total volume flow \cite{Kamsma2023UnveilingIontronics}. Here we note that the flow is incompressible such that the total flow $Q(V)$ is a spatial constant that therefore depends on an effective surface potential denoted by $\overline{\psi}_0$, which in turn depends via the Gouy-Chapman relation on the total salt concentration through the Debye length. Therefore Eq.~(\ref{eq:rhos}) in its full glory is actually a self-consistency problem that requires a ``closure'' to obtain $\overline{\psi}_0$ from a spatially varying zeta potential $\psi_0(x)$ or surface charge $-e\sigma(x)$. In principle such a problem can be solved through an iterative scheme, which would for instance be of interest for specific investigations of the volume flow characteristics (in fact our finite-element approach of the PNPS equations coupled to the Langmuir dynamics as presented in the main text solves the strongly coupled full set of equations numerically). Here, however, we seek some further insight from a simple AA and hence define $\overline{\psi}_0$ as the Gouy-Chapman potential that corresponds to the (time-dependent) average surface charge density $\overline{\sigma}(t)=\int_0^L R(x)\sigma(x,t)\dd x/\int_0^L R(x)\dd x$ at the equilibrium (i.e.\ $\overline{\rho}_s(x,V)=\sum_{i}\rho_{i,\mathrm{b}}$) salt concentration.

Since the salt does not accumulate uniformly inside the channel, see Eq.(\ref{eq:rhos}), the surface in different regions in the channel charge differently. A convenient proxy for the resulting inhomogeneous surface charge is coarsely captured by the typical difference $\sigma^{\prime}(t)=\sigma(0.9L,t)-\sigma(0,t)$ of the surface charge density over the channel. However, for our system parameters we found that the effect of including a finite $\sigma^{\prime}(t)$ is relatively small compared to the approximation of a homogeneous surface charge. Hence the essence of  understanding the underlying mechanism can also be extracted from an AA that incorporates salt accumulation in a conical channel with a homogeneous surface charge \cite{Boon2022Pressure-sensitiveGeometry}. In the inset of Fig.~2(b) in the main text, we plot $\partial_t\overline{\sigma}$ as a function of $\overline{\sigma}$, instead of treating $\sigma(x)$ an independent output across the channel, so in this instance we simply used $\sigma(x,t)=\overline{\sigma}(t)$ for all $x$.
\begin{figure*}[ht!]
\centering
     \includegraphics[width=1\textwidth]{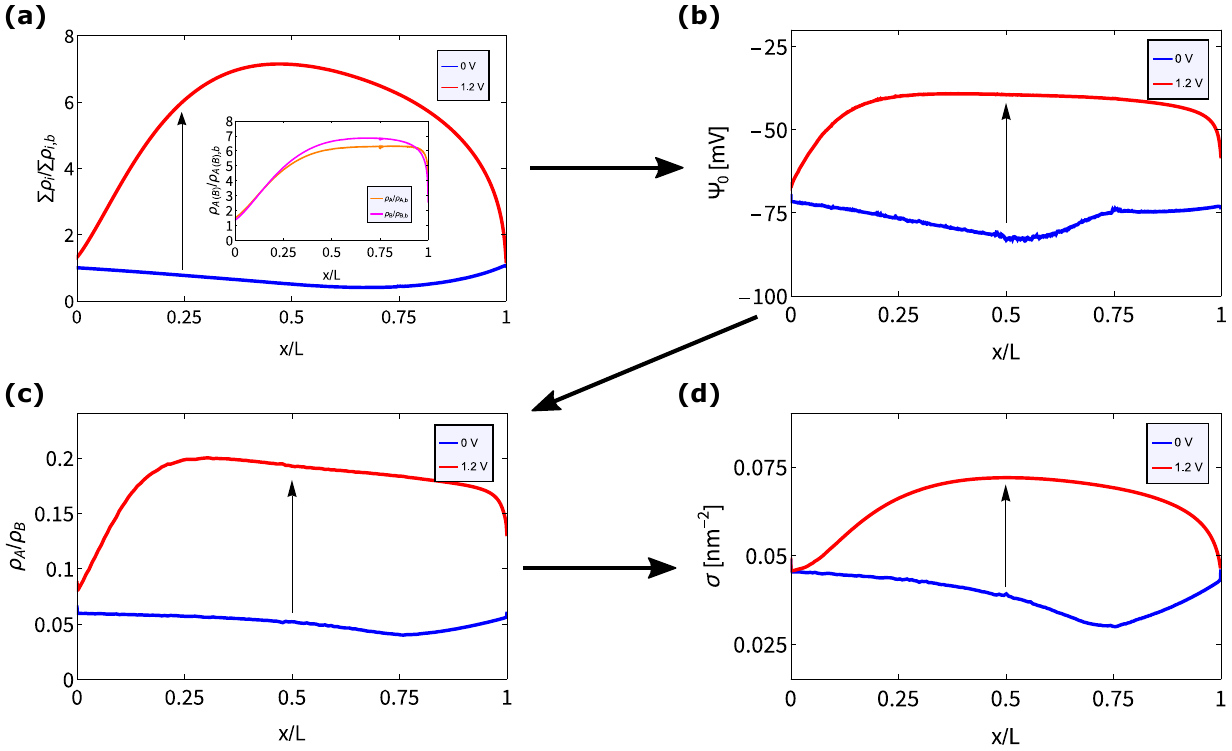}
        \caption{FE results before (blue) and during (red) the voltage pulse of 1.2 V, displaying the mechanism as described by our AA that underpins the fast charging and subsequent slow discharging. \textbf{(a)} The total salt concentration $\sum_i\rho_i/\sum_i\rho_{i,\ch{b}}$ at the channel's central axis shortly after the 1.2 V pulse is applied, showing strong salt concentration polarisation by up to a factor ~7. As shown in the inset, the central axis concentrations of the reactant ions increase nearly identically, which demonstrates that the salt concentration polarisation is not responsible for the change in concentration ratio at the surface. \textbf{(b)} The surface potential $\psi_0$ shortly after the voltage is applied but before the surface has had time to significantly charge. Due to the increased salt concentration in the channel, the surface charge is better screened and hence the surface potential $\psi_0$ becomes significantly less negative after the pulse by several tens of mV. \textbf{(c)} The concentration ratio $\rho_A/\rho_B$ of the reactant ions at the channel wall increases by up to a factor 4 due to the less-negative surface potential, in line with the AA prediction that $\rho_{\mathrm{A}}/\rho_{\mathrm{B}}=\exp[e\psi_{0}(x)/k_{\mathrm{B}}T]$. \textbf{(d)} The surface density profile $\sigma(x)$ before and at the very end of the voltage pulse, showing a considerable increase in $\sigma$ due to the relative increase of A$^{-3}$ over B$^{2-}$.}
        \label{fig:AAfigSM}
\end{figure*} 

Clearly, the backward and forward surface reaction rates of the Langmuir kinetics described by Eq.~(2) of the main text require the reactant ion concentrations $\rho_{\mathrm{A}}$ and $\rho_{\mathrm{B}}$ at the channel wall. In the AA, we assume that both ion species A$^{3-}$ and B$^{2-}$ satisfy a Boltzmann distribution such that the concentration of A$^{3-}$ at the channel wall at radial distance $r=R(x)$ from the axis is for $x\in[0,L]$ given by
\begin{align}\label{eq:rhoAB}
    \rho_{\mathrm{A}}(x,R(x),t)=\frac{\rho_{\mathrm{A,b}}}{2\rho_{\pm,\mathrm{b}}}\overline{\rho}_s(x,V(t)) e^{-z_{\mathrm{A}}e\psi_0(x)/k_{\mathrm{B}}T},
\end{align}
and likewise for $\rho_{\mathrm{B}}(x,R(x),t)$ of species B$^{2-}$.
Here $\rho_{\mathrm{A,b}}/2\rho_{\pm,\mathrm{b}}$ is the ratio of the bulk concentrations of $\text{A}^{3-}$ ions with valency $z_A=-3$ (or $\text{B}^{2-}$ ions with valency $z_B=-2$) and the monovalent background ions, $\psi_0(x)=-\left(2k_{\mathrm{B}}T/e\right)\text{sinh}^{-1}\left(2\pi\lambda_{\mathrm{B}}\lambda_{\mathrm{D}}(x)\sigma(x)\right)$ the Gouy-Chapman surface potential at surface charge density $-e\sigma(x)$, $\lambda_{\mathrm{D}}(x)=1/\sqrt{4\pi\lambda_B\overline{\rho}_{\mathrm{s}}(x,V(t))}$ the (local) Debye length, and $\lambda_{\mathrm{B}}=e^2/4\pi\epsilon k_{\mathrm{B}}T\approx0.7\text{ nm}$ the Bjerrum length. The implicit assumption of Eq.(\ref{eq:rhoAB}) is that the accumulation factor of the two reactive ion species on the central axis is the same as that of the monovalent background ions.

Eq.~\ref{eq:rhoAB} is quantitatively a rather coarse approximation because Eq.~(\ref{eq:rhos}) was originally derived in the case of a pure 1:1 electrolyte in conical channels with $\lambda_{\mathrm{D}}\ll R(x)$ \cite{Kamsma2023UnveilingIontronics,Boon2022Pressure-sensitiveGeometry}. Neither of these two assumptions are satisfied for the parameters in the main text. However, since $\rho_{\mathrm{A,b}},\rho_{\mathrm{B,b}}\ll\rho_{\pm,\mathrm{b}}$ the 1:1 background electrolyte makes up for the vast majority of ions in the channel, and for most of the channel $\lambda_{\mathrm{D}}\ll R(x)$ is still reasonably satisfied. Therefore, as we presented in the main text, this approach is still sufficient to qualitatively unveil (at least part of) the mechanism behind the fast charging and slow discharging processes that facilitate the LTP, while the quantitative differences can be explained by these coarse assumptions.

The timescale $\tau=L^2/12D\approx 5.6$ ms for voltage-driven concentration polarisation for the system parameters of the main text \cite{Kamsma2023UnveilingIontronics,Kamsma2023IontronicMemristors} is much faster than the (dis)charging timescales set by the reaction rate $k$ and the (typical) concentrations of the reactive ions. In this reaction-limited rather than diffusion-limited regime, we can safely assume that the time-dependent total ion concentration $\overline{\rho}_s(x,V(t),t)=\overline{\rho}_s(x,V(t))$, such that the we only need to solve the temporal dynamics for the surface charge as described by the Langmuir kinetics
\begin{align}\label{eq:langmuirdisplacementAA}
    \dfrac{\partial \sigma}{\partial t}=k\Big(\rho_{\ch{A}}(\Gamma-\sigma)-K\rho_{\ch{B}}\sigma\Big),
\end{align}
with $\rho_{\ch{A}}=\rho_{\mathrm{A}}(x,R(x),t)$ as given by Eq.~(\ref{eq:rhoAB}) (likewise for ion B$^{2-}$). Therefore, Eq.~(\ref{eq:langmuirdisplacementAA}) can quickly and easily be (numerically) solved for $\sigma(x,t)$ across the channel length to obtain the results for $\overline{\sigma}(t)$ and $\partial_t\overline{\sigma}(t)$ presented in Fig.~2(b) of the main text.
\begin{figure*}[ht!]
\centering
     \includegraphics[width=1\textwidth]{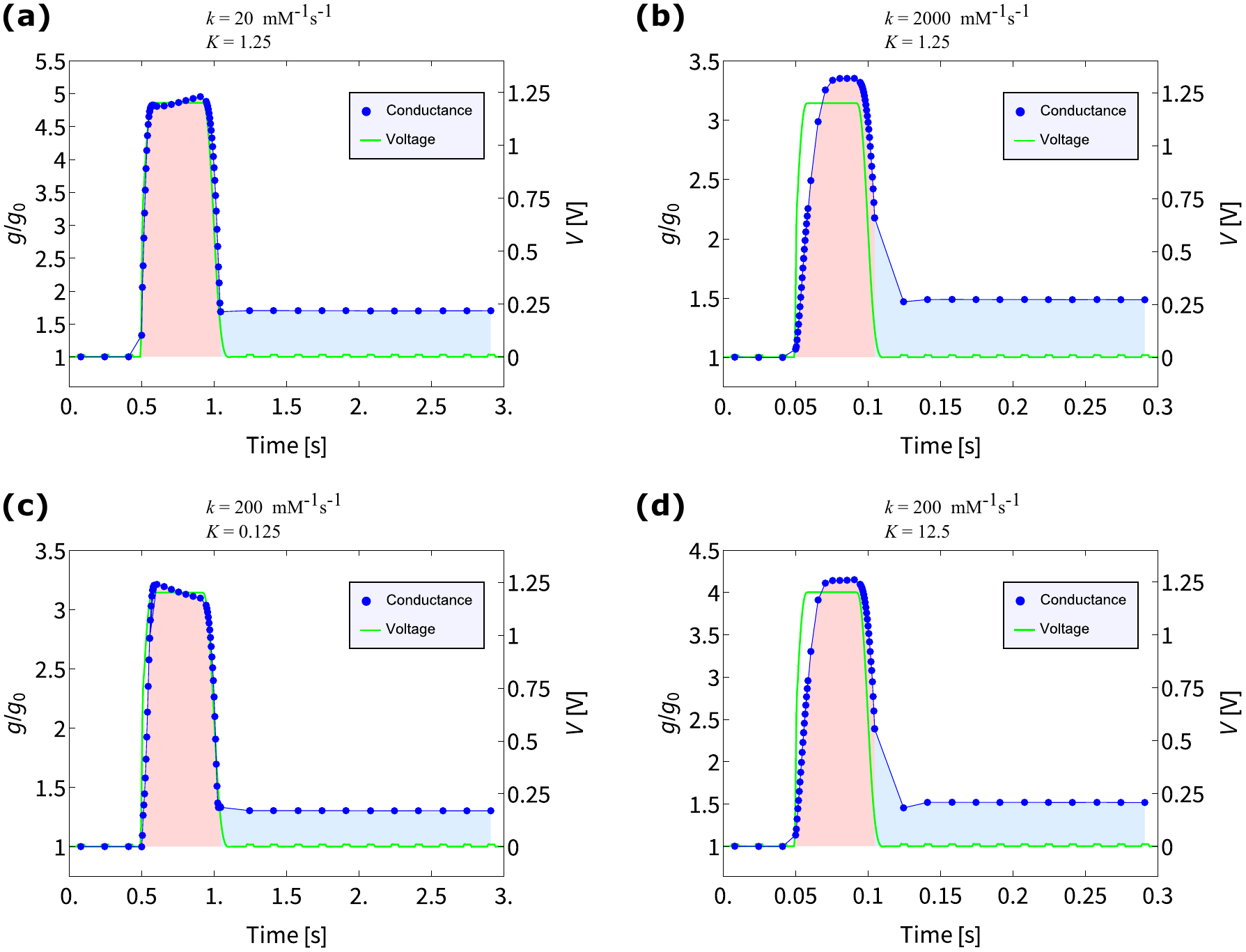}
        \caption{Near identical fast charging and slow discharging dynamics for the reaction parameters \textbf{(a)} $k=20\text{\,mM}^{-1}\text{s}^{-1}$ and $K=1.25$, \textbf{(b)} $k=2000\text{\,mM}^{-1}\text{s}^{-1}$ and $K=1.25$, \textbf{(c)} $k=200\text{\,mM}^{-1}\text{s}^{-1}$ and $K=0.125$, and \textbf{(d)} $k=200\text{\,mM}^{-1}\text{s}^{-1}$ and $K=12.5$. For Fig.~\ref{fig:AlternParams}(a) the concentrations at the reservoir $\rho_{\ch{A,b}}$ and $\rho_{\ch{B,b}}$ were both increased to 0.2 mM, for Fig.~\ref{fig:AlternParams}(b) the voltage pulse was shortened to 50 ms, for Fig.~\ref{fig:AlternParams}(c) no adjustments had to be made, and for Fig.~\ref{fig:AlternParams}(d) only the reservoir concentrations $\rho_{\ch{A,b}}$ of $\text{A}^{3-}$ were increased to 0.2 mM. These concentrations and voltage pulse durations are experimentally facile to control.}
        \label{fig:AlternParams}
\end{figure*} 
The above equations describe how an applied positive voltage creates salt accumulation described by $\overline{\rho}_s(x,V)$ (Eq.~(\ref{eq:rhos})), which renders the (Gouy-Chapman) surface potential  $\psi_0(x)=-\left(2k_{\mathrm{B}}T/e\right)\text{sinh}^{-1}\left(2\pi\lambda_{\mathrm{B}}\lambda_{\mathrm{D}}(x)\sigma(x)\right)$ less negative, which in turn increases the ratio $\rho_{\mathrm{A}}/\rho_{\mathrm{B}}=\exp[e\psi_{0}(x)/k_{\mathrm{B}}T]$ (Eq.~(\ref{eq:rhoAB})) towards a higher concentration of $\text{A}^{3-}$, and consequently charges the surface to a deeper negative charge density  as described by $\partial_t\sigma$ (Eq.~(\ref{eq:langmuirdisplacementAA})). After the voltage is removed and the surface is relatively deeply negatively charged, not only the concentration of ions on the central axis quickly decreases (on the time scale $\tau$) to undo the earlier accumulation (Eq.~(\ref{eq:rhos})), but also the deeply negative surface charge strongly repels the negative reactive ions away from the surface according to Eq.~(\ref{eq:rhoAB}).  Both effects contribute to significantly lower ion concentrations at the surface, leading to a much slower discharging process compared to the initial charging process (Eq.~(\ref{eq:langmuirdisplacementAA})).

\section{Finite-element demonstration of mechanism}
In Fig.~\ref{fig:AAfigSM} we show profiles between base (at $x=0$) and tip (at $x=L$)  as obtained from the finite-element calculations at times before (blue) and after the applied voltage of 1.2 V is switched on (red), that demonstrate and confirm the mechanism derived in Sec.~\ref{sec:AA}. The profiles for the case before the applied voltage already show spatial inhomogeneities. These stem from (weakly) overlapping double layers towards the tip of the channel that create a nonzero steady-state electric (``Donnan'') potential profile on the central axis within the channel, even without an applied external voltage, while charge neutrality in each slab remains ensured. This internal electric potential profile slightly shifts the salt concentration in steady-state in the channel (and thereby the surface potential and charge) and is therefore responsible for these observed inhomogeneous profiles before the applied voltage.

\begin{figure*}[ht!]
\centering
     \includegraphics[width=1\textwidth]{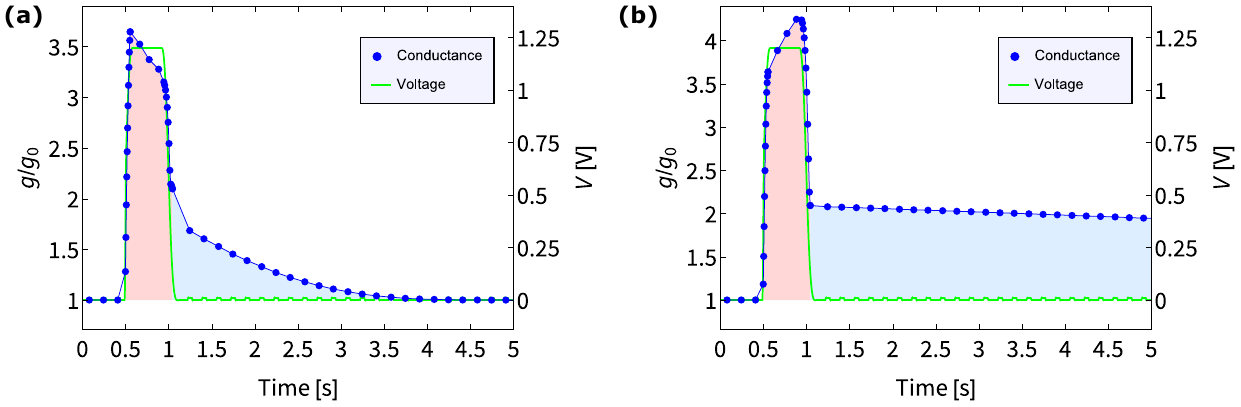}
        \caption{Channel conductance $g(t)$ for different reaction mechanism upon a 1.2 V voltage pulse for $t\in[0.5,1]$ s. \textbf{(a)} Change in conductance for a channel featuring a simple adsorption reaction as in Eq.~(\ref{eq:adsorption}), featuring a conductance modulation that lasts much longer than the concentration polarisation timescale $\tau\approx 5.6$ ms, but much shorter than the displacement reactions. \textbf{(b)} Change in conductance for a channel featuring a 2:1 displacement reaction as in Eq.~(\ref{eq:21displacement}). The increase in conductance remains for much longer than the signal duration, similar to the 3:2 displacement reaction in the main text.}
        \label{fig:LTP_SMfig}
\end{figure*} 
Fig.~\ref{fig:AAfigSM}(a) shows the profile of the total ion concentration relative to the bulk concentrations, clearly revealing that the applied voltage induces salt accumulation by a factor as large as $\sim7$, where the inset shows that A$^{3-}$ and B$^{2-}$ ions indeed both show similar accumulation on the central axis, supporting our approach of Eq.~(\ref{eq:rhoAB}). Due to enhanced screening by the salt accumulation, the surface charge gets better screened during the applied voltage, which renders the surface potential $\psi_0(x)$ less negative immediately (a few $\tau$) after the voltage is applied, as we show in Fig.~\ref{fig:AAfigSM}(b). Due the valency difference between the ions A$^{3-}$ and B$^{2-}$, the less-negative surface potential increases the ratio $\rho_{\mathrm{A}}/\rho_{\mathrm{B}}$ at the channel wall, as shown in Fig.~\ref{fig:AAfigSM}(c). This, in turn,  shifts the displacement reaction balance, driving more A$^{3-}$ to bind to the functionalized surface groups and thereby increasing $\sigma(x)$, i.e.\ rendering the surface charge $-e\sigma(x)$ substantially more negative just before the end of the pulse, as shown in Fig.~\ref{fig:AAfigSM}(d).

Lastly, we remark that the PNPS equations are less accurate for electrolytes with ions of valence $|z_i|>1$ \cite{Ebeling2011ElectrostaticValency}. However, since $\rho_{\mathrm{A,b}},\rho_{\mathrm{B,b}}\ll\rho_{\pm,\mathrm{b}}$ the electric potential profile $\Psi(x,r,t)$, the electro-osmotic flow $\mathbf{u}(x,r,t)$ and the ionic fluxes $\mathbf{j}_{i}(x,r,t)$ are dominated by the 1:1 electrolyte component.

\section{Alternative reaction parameters}
The results in the main text (mostly) focused on a single set of reaction parameters with a reaction rate of $k=200\text{\,mM}^{-1}\text{s}^{-1}$ and equilibrium constant $K=1.25$. However, based on our above explanation of the fast (slow) (dis)charging, the mechanism that underpins this contrast in timescales is based on the combined effect of (i) an increased charging rate due to ion accumulation during the voltage pulse and (ii) a decreased discharging rate caused by relatively low concentrations of the multivalent reactants after the pulse due to their enhanced Coulomb repulsion from the persistently high surface charge. Therefore, we expect that this qualitative feature is mostly independent from specific reaction parameters.

To demonstrate this parameter insensitivity, we performed additional FE calculations where we changed $k$ and $K$ by an order of magnitude compared to the parameters in the main text. The results are shown in Fig.~\ref{fig:AlternParams}, where we see near-identical fast-slow dynamics for (a) $k=20\text{\,mM}^{-1}\text{s}^{-1}$ and $K=1.25$,  
(b) $k=2000\text{\,mM}^{-1}\text{s}^{-1}$ and $K=1.25$, 
(c) $k=200\text{\,mM}^{-1}\text{s}^{-1}$ and $K=0.125$, 
and (d) $k=200\text{\,mM}^{-1}\text{s}^{-1}$ and $K=12.5$
.

Although the relative ratio of the fast charging and slow discharging time scale is largely independent from $k$ and $K$, these do influence the absolute overall timescales. To adjust for this, and to illustrate the versatility, tunability, and robustness of the system, we tuned two more parameters that are experimentally easily accessible, namely the equilibrium concentrations of the reactant ions or the voltage pulse duration. Choices for these parameters are guided by the rough constraint provided in the main text. For Fig.~\ref{fig:AlternParams}(a) the concentrations at the reservoir $\rho_{\ch{A,b}}$ and $\rho_{\ch{B,b}}$ were both increased to 0.2 mM, for Fig.~\ref{fig:AlternParams}(b) the voltage pulse was shortened to 50 ms, for Fig.~\ref{fig:AlternParams}(c) no further adjustments were made, and for Fig.~\ref{fig:AlternParams}(d) only the reservoir concentration $\rho_{\ch{A,b}}$ was increased to 0.2 mM.

\section{Alternative reaction mechanisms}
In the main text we used a 3:2 substitution reaction mechanism for our results. To show that (at least qualitatively) similar LTP features can be extracted using other reaction mechanisms we consider a simple adsorption reaction of the form
\begin{align}\label{eq:adsorption}
    \ch{S}_{\ch{(s)}}+\ch{A}_{\ch{(aq)}}^{-}\xrightleftharpoons[k_{\ch{des}}]{k_{\ch{ads}}\rho_{\ch{A}}}\ch{SA}_{\ch{(s)}}^{-},
\end{align}
where the surface charge density $-e\sigma(x,t)$ is now described by Langmuir kinetics of the form
\begin{align}\label{eq:adsorptionLangmuir}
    \dfrac{\partial \sigma}{\partial t}=k_{\ch{ads}}\rho_{\ch{A}}(\Gamma-\sigma)-k_{\ch{des}}\sigma,
\end{align}
with $k_{\ch{ads}}=10\text{ mM}^{-1}\text{s}^{-1}$, $k_{\ch{des}}=0.4\text{ s}^{-1}$, and $\rho_{\mathrm{A,b}}=0.02$ mM.

We also investigate a 2:1 displacement reaction mechanism of the form
\begin{align}\label{eq:21displacement}
    \ch{SB}_{\ch{(s)}}+\ch{A}_{\ch{(aq)}}^{2-}\xrightleftharpoons[kK\rho_{\ch{B}}]{k\rho_{\ch{A}}}\ch{SA}_{\ch{(s)}}^{-}+\ch{B}_{\ch{(aq)}}^{-},
\end{align}
described by the same Langmuir kinetics as in the main text
\begin{align}\label{eq:21displacementlangmuir}
    \dfrac{\partial \sigma}{\partial t}=k\Big(\rho_{\ch{A}}(\Gamma-\sigma)-K\rho_{\ch{B}}\sigma\Big),
\end{align}
with parameters as in the main text.

In Fig.~\ref{fig:LTP_SMfig}(a) we show the results for the adsorption reaction of  Eq.~\ref{eq:adsorption}. It is clear that quantitatively the memory effect is significantly shorter, suggesting that for neuromorphic applications this reaction mechanism is less interesting and less relevant. However, the increase of the conductance still lasts several seconds, i.e.\ orders of magnitude longer than the typical diffusion memory timescale $\tau\approx 5.6$ ms for these parameters. Moreover, the fact that the conductance dynamics is apparently different for different substitution mechanism could make it more relevant for sensing purposes, giving a clearly identifiable fingerprint for distinct underlying reaction mechanisms. In Fig.~\ref{fig:LTP_SMfig}(b) we show the channel conductance for the 2:1 displacement reaction of Eq.~\ref{eq:21displacement}. Here we clearly see that the conductance is retained over a much longer timescale than the input signal. Although quantitatively the fast charging and slow discharging is not quite as extreme as the 3:2 displacement reaction from the main text, it is still a very notable separation of charging and discharging timescales, demonstrating that this reaction mechanism could provide an alternative for the 3:2 displacement reaction mechanism considered in the main text.

\begin{figure}[t]
\centering
     \includegraphics[width=0.49\textwidth]{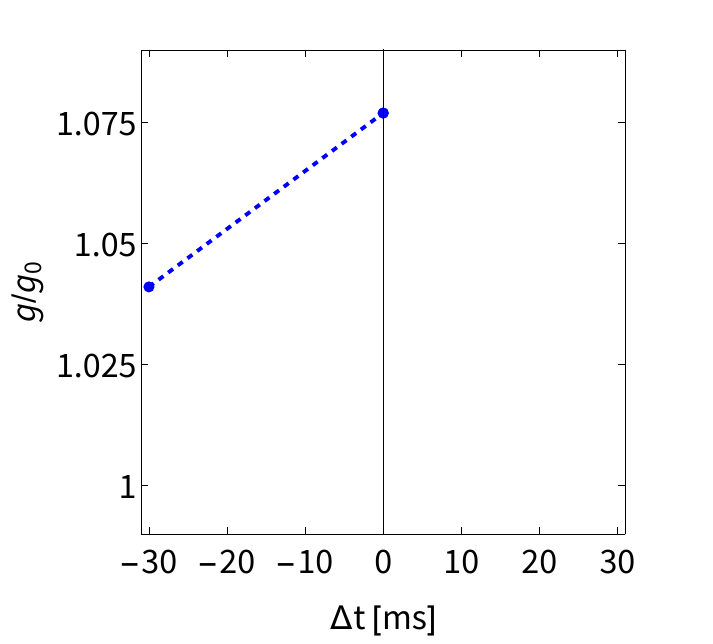}
        \caption{Diminishing potentiation for an FE model with a 30 ms lasting flux injection of $\text{A}^{3-}$ at $\sim5\text{ }\mu$m away from the channel base, while all concentration are kept constant at the boundary $\sim10\text{ }\mu$m away from the ion flux injection point. Although parameters are not optimized for this method of modeling a chemical signal, a clear diminishing potentiation effect is again visible.}
        \label{fig:STDPflux}
\end{figure} 
\section{Chemical signal with flux injection}
In the main text, the simple yet effective method of a time-dependent boundary condition $\sim10\text{ }\mu$m away from the base was chosen to model a concentration shock. This effectively captures the two relevant features that arise from an ion injection, namely, (i) a diffusive influx from an injection location to the base of the channel to increase the local concentration, and (ii) a diffusive dissipation into the rest of the large reservoir after the injection stops. Raising the concentration of $\text{A}^{3-}$ at the boundary results in a diffusive influx of these ions as long as it is below the target concentration. Therefore this is equivalent with an ion injection. The ion coincidence detection (Fig.~4(a)) and the depression side of the STDP (Fig.~4(b, red)) do not rely on how (or whether) these ions subsequently dissipate into the rest of the base reservoir, as we showed in Fig.~4(a,magenta) that the presence of extra ions without any voltage has no significant effect. However, the potentiation side of the STDP (Fig.~4(b, blue)) relies on the dissipation of $\text{A}^{3-}$ into the rest of the reservoir for the diminishing potentiation effect. It might have been more accurate to inject the ions near the base of the channel and simultaneously keep the boundary concentration fixed further away. Rather, for simplicity, in the main text we model the end of the shock by reverting the concentration at the same place as the injection. This possibly speeds up the dissipation process somewhat because in reality the boundary condition with the original lower concentration should be further into the reservoir from the point of injection. However, otherwise the STDP results in Fig.~4 do not rely on this choice of boundary conditions.

To demonstrate that the spike time-dependent potentiation indeed also occurs for a fixed boundary concentration combined with an ion injection near the base, we calculated two data points on the potentiation side of the STDP curve with an altered FE model. Here, an inward $\text{A}^{3-}$ flux of $5\cdot10^{-3}$ mol/(m$^2$s) that lasts 30 ms is placed at a circular patch (placed on the surface of the membrane) of height 1 $\mu$m revolved around the central axis, roughly 5 $\mu$m away from the channel base. During the 30 ms ion injection, $\rho_+$ is also injected to maintain charge neutrality. All concentrations are now kept constant at the reservoir boundary $\sim10\text{ }\mu$m away from the ion flux injection point. The results are shown in Fig.~\ref{fig:STDPflux}. Although the parameters are not fully optimized for this method of modeling a chemical signal, a clear diminishing potentiation effect is again visible. Moreover, the time window for this potentiation is still on the order of 10 ms. Therefore the boundary condition implementation in the main text is a coarse, but effective method, that simplifies the modeling while not interfering with the relevant results presented.

%